\documentclass[utf8]{FrontiersinHarvard} % for articles in journals using the Harvard Referencing Style (Author-Date), for Frontiers Reference Styles by Journal: https://zendesk.frontiersin.org/hc/en-us/articles/360017860337-Frontiers-Reference-Styles-by-Journal
\usepackage{url,hyperref,lineno,subcaption}
\usepackage[onehalfspacing]{setspace}
\def\keyFont{\fontsize{8}{11}\helveticabold }
\def\firstAuthorLast{Rodríguez-Ardila {et~al.}} %use et al only if is more than 1 author
\def\Authors{Alberto Rodr\'{\i}guez-Ardila\,$^{*,1,2,3}$ and Fernando Cerqueira-Campos\,$^{2}$}
\def\Address{$^{1}$Laboratory Nacional de Astrof\'{\i}sica, Rua dos Estados Unidos, 154. CEP 37504-364, Itajubá, MG, Brazil \\
$^{2}$Instituto Nacional de Pesquisas Espaciais, Avenida dos Astronautas 1758 - Jardim da Granja S\~ao Jos\'e dos Campos/SP - CEP 12227-010, Brazil\\
$^{3}$Observatório Nacional, Rua General José Cristino 77, CEP 20921-400, São
Cristóvão, Rio de Janeiro, RJ, Brazil}
\def\corrAuthor{Alberto Rodríguez-Ardila}

\def\corrEmail{aardila@lna.br}

\begin{document}
\onecolumn
\firstpage{1}

\title[The CLR in AGN]{The Coronal Line Region of Active Galactic Nuclei} 

\author[\firstAuthorLast ]{\Authors} %This field will be automatically populated
\address{} %This field will be automatically populated
\correspondance{} %This field will be automatically populated

\extraAuth{}% If there are more than 1 corresponding author, comment this line and uncomment the next one.
%\extraAuth{corresponding Author2 \\ Laboratory X2, Institute X2, Department X2, Organization X2, Street X2, City X2 , State XX2 (only USA, Canada and Australia), Zip Code2, X2 Country X2, email2@uni2.edu}

\maketitle

\begin{abstract}

%%% Leave the Abstract empty if your article does not require one, please see the Summary Table for full details.
\section{}
Forbidden coronal lines has traditionally called the attention due to the high-energy photons required for their production (IP $>$100~eV, where IP is the ionisation potential of the transitions that originate the line). As such, they are regarded as the most highly ionised component of Active Galactic Nuclei (AGN). For decades, it was thought that they were only formed in the inner portions of the narrow line region (NLR). Nowadays, due to the larger sensitivity of the detectors and the availability of integral field unit (IFU) spectrographs, that emission in addition to the nuclear component is found to be extended up to a few kiloparsecs away from the active centre. In this review, we highlight the most important aspects of the coronal emission and discuss the recent developments in the field. In particular, we emphasize the discovery that they can be used to determine the mass of the central supermassive black hole, to reconstruct the SED, as well as to trace the most energetic feedback component of the ionised gas in AGN.

\tiny
 \keyFont{ \section{Keywords:} line:formation -- galaxies: active -- galaxies: jets -- galaxies: Seyfert -- line: profiles} %All article types: you may provide up to 8 keywords; at least 5 are mandatory.
\end{abstract}

%\section{Introduction}

%For Original Research Articles %\citep{conference}, Clinical Trial Articles %\citep{article}, and Technology Reports %\citep{patent}, the introduction should be %succinct, with no subheadings \citep{book}. For %Case Reports the Introduction should include %symptoms at presentation \citep{chapter}, physical exams and lab results \citep{dataset}.

\section{Introduction}

Coronal lines (CLs), or high-ionisation emission lines, originate from forbidden fine-structure transitions excited through collisions in highly ionized species (ionization potential, IP $>$ 100 eV). For this reason they are considered a 
reliable signature of the presence of an AGN in galaxies \citep{penston+84,oliva_1994,ardila_2002,komossa+08}. We notice, though, that CLs
are also detected in spectra of supernova remnants \citep{oliva_1999,smith+09}, planetary nebulae \citep{pottasch+09} and Wolf–Rayet stars
\citep{schaerer+99}. However, the typical luminosities in these latter three
classes of objects are low, $\sim$10$^{31-33}$~erg\,cm$^{-2}$\,s$^{-1}$ while in the former the luminosity amounts to $\sim$10$^{38}$~erg\,cm$^{-2}$\,s$^{-1}$ at the very least. Therefore, it would be necessary tens of thousands of sources of stellar nature to produce detectable CLs in AGN if the central engine is discarded
as their origin.  In this review we will focus on the coronal emission due to AGN activity. 

Observationally, the most common optical CLs in AGN are [Ne\,{\sc v}]$\lambda$3425, [Fe\,{\sc vii}]$\lambda$6087, [Fe\,{\sc x}]$\lambda6374$, and [Fe\,{\sc xi}]$\lambda7889$ \citep{mazzalay_2010}. Part of these lines had already been identified in the original work
of \citet{seyfert43}. \citet{olimor90} increased the interest in studying CLs after the detection of [Si\,{\sc vi}]~1.963$\mu$m in a the near-infrared region (NIR). They identified that line as a genuine tracer of AGN activity. Later, detections of [S\,{\sc ix}]~1.252$\mu$m, [Si\,{\sc x}]~1.421$\mu$m, [S\,{\sc xi}]~1.921$\mu$m,  [Ca\,{\sc viii}]~2.321$\mu$m, and [Si\,{\sc vii}]~2.483$\mu$m  
\citep{thompson96}, [Al\,{\sc ix}]~2.048$\mu$m \citep{maiolino97}, [Mg\,{\sc viii}]~3.027$\mu$m, [Mg\,{\sc vii}]~5.503, 9.030$\mu$m, [Mg\,{\sc v}]~5.601$\mu$m, [Ne\,{\sc vi}]~7,652$\mu$m, and [Ne\,{\sc v}]~14.322$\mu$m \citep{moorwood_1996}, and [Fe\,{\sc xiii}]~1.074,1.078~$\mu$m \citep{ardila_2002} appeared in samples of AGN in the optical, near- and mid-infrared spectra of such objects.

The mechanisms responsible for the production of the high ionization state of these
elements is still not very clear. Basically, two processes have been proposed. The first one is photoionization by the central source, where due to the intense AGN continuum, particularly in far-UV and soft x-rays \citep{shields+75,grandi78,ferguson1997},  strongly ionise the gas allowing the production of such lines. The second alternative is shocks, where the material ejected by the AGN or a jet mechanically interacts with the gas of the circumnuclear region \citep{osterbrock+64,oke+68,dopita+95} giving rise to those lines. There is yet a third, intermediate possibility, which is the combination of the above two processes\citep{viegas+89,olimor90,contini+01,ardila_2006}.

Observations carried out in the 80’s in the optical region showed that the centroids of 
CLs are usually displaced to the
blue with respect to the systemic velocity of the galaxy. Furthermore, CLs
tend to be broader than low ionization forbidden lines. Other studies indicated a correlation
between the width of these lines and the ionization potential required to ionize the
material \citep{pelat+81,evans88}  and the critical density of the corresponding transition. The interpretation of
these results was that the coronal line region (CLR), the region where these lines are formed, is located between the NLR and the broad line region (BLR).   However, this scenario 
was later challenged in several works, initially by \citet{prieto_2005},
and later by \citet{ardila_2006} and \citet{muller_2011}. They all showed that the region emitting the coronal lines is extended on scales of hundreds
of parsecs, including a very compact, seeing-limited component. 

Although CLs are ubiquitous in the spectra of AGNs, not all AGNs display them.
\citet{ardila_2011} using a sample of 47 AGNs observed in the NIR
found that in 67\% of objects, a CL is identified. They
showed that the lack of CLs in the remaining 33\% of objects is genuine and is not motivated by
detection problems due to sensitivity.  
Similar results regarding the frequency of CLs were also found by \citet{lamperti+17}. In the optical region, CLs of neon ([Ne\,{\sc v}]) and iron ([Fe\,{\sc vii}] and [Fe\,{\sc x}]) are usually observed but quantitative studies of their frequency are scarce. This is probably because they tend to be faint when compared to other NLR lines. Overall, [Ne\,{\sc v}]$\lambda$3425 is the brightest CL followed by [Fe\,{\sc vii}]~$\lambda$6087.  They both can reach EW $> 10$~\AA\ for dustless high Z gas \citep{mckaig+24}. Observationally, the former line can be as bright as H$\beta$ while the latter may display up to 25\% of that H\,{\sc i} line \citep{mazzalay_2010, rose+15a}.

The importance of studying of the coronal emission has varied with time. Up to a decade ago, most works focused on the mechanisms responsible for their production as well as to determine the precise location where they were formed. The discovery that the coronal emission, in addition to the unresolved nuclear component, is extended to scales of a few kiloparsecs \citep{ardila_2020,negus_2021}, and that this emission traces effectively the highest ionised portion of the gas participating in the feedback process \citep{muller_2011, mazzalay_2013,ardila_2017_ngc_1386} attracted the interest of the AGN community. Another important development was the discovery that these lines can be used to determine the mass of the central black hole \citep{prieto+22} using single-epoch spectra. The fact that [Ne\,{\sc v}]~$\lambda$3425 is the bluest and brightest coronal line in the optical region makes it suitable to find out AGN up to $z \sim$1.4 in optical surveys (Gulli et al. 2010) or in faint galaxies at high-redshifts (Li et al. 2024; Chisholm et al. 2024).

For all the above, in this review we highlight the most important aspects of the CLs (Sect.~\ref{sec:clr}), discuss the drivers for studying this emission  (Sect.~\ref{sec:importance}) as well as comment on open issues that are critical for its understanding (Sect.~\ref{sec:open}). Final remarks are in Sect.~\ref{sec:remarks}.

\section{The Coronal Line Region} 
\label{sec:clr}

Until the early 1990's, the general consensus was that the coronal lines were emitted in very compact, unresolved region, between the BLR and the NLR \citep{penston+84}. Dust at the inner edge of the torus would be evaporated by the strong radiation field of the AGN, producing the high-ionisation lines observed in AGN spectra \citep{pier_1994}. Later, \citet{murayama_1998} expanded this paradigm through the study of the [Fe\,{\sc vii}]~$\lambda$6087 in a sizeable sample of Type~I and~II AGNs. The fact that the former sources typically display broader [Fe\,{\sc vii}] lines than those observed in the latter led them to proposed that the CLs would form in three different regions:  (i) the inner wall of the dusty torus, (ii) clouds associated with the narrow-line region at $\sim$10 to $\sim$100 pc from the AGN, and (iii) the extended ionized region at $\sim$1 kpc from the central engine. However, it is important to notice that this proposition was drawn from integrated long-slit spectra and based on the differences in line width between the two main types of AGN. 

In the 2000s, optical studies using the HST (Hubble Space Telescope) and NIR observations,
using ground-based telescopes with adaptive optics, sought to determine the size and
CLR morphology of local AGNs \citep{riffel_2008,storchi-bergmann_2009,mazzalay_2010,muller_2011,muller+18}. What
these authors identified was that the CL emission region presents complex and 
extended morphologies depending of the CL  studied with
strong differences in their intensities and line profiles  throughout a spatial mapping.
Moreover, in most of the cases when extended emission was observed, it is closely aligned with the position axis of the radio jet.
All these works found that CLs are emitted from a very compact region (located between the BLR
and the NLR, with sizes of approximately 30 pc) to more extended regions, reaching around 250 pc from the nucleus \citep{murayama_1998,muller_2011,muller+18}. It is important to notice that due to the small field-of-view of the detectors, it is likely that the CLR is still more extended. This was indeed discovered later on (see below).

The hypothesis of a compact component of the CLR was later reinforced by variability detected in the [Fe\,{\sc vii}]~$\lambda6087$ CL of NGC\,4151 \citep{landt_4151_15a} and NGC\,5548 \citep{landt_5548_15b,kynoch+22}. These results imply that at least part of the flux measured in CLs under seeing-limited conditions arise in the innermost portion of the NLR, likely photoionised by radiation from the central engine. Still, temporal changes in the CL of [Fe\,{\sc x}]~$\lambda$6374 and the broad Balmer lines, accompanied by substantial changes in the shape of the optical and X-ray continua in NGC\,1566  \citep{oknyansky+19} led these authors to propose the changing-look nature of that AGN. However, it was just until very recently that the the size of the compact portion of the CLR could be first determined. GRAVITY \citep{gravity_2021_NGC3783} provided the first opportunity to measure the size of the nuclear CLR through the [Ca\,{\sc viii}] line. They found a emitting CLR of 0.4~pc, firmly placing it at the very inner regions of the NLR, outside the BLR and the inner face of the torus. Combined with VLT/SINFONI data, that work also showed that the CLR is composed of a bright compact nuclear component and a fainter extended component out to 100~pc with outflow
kinematics.

\citet{muller_2011,mazzalay_2013,ardila_2017_ngc_1386,may_2018,fonseca-faria_2021} and \citet{Speranza2022}  expanded the understanding of the CLR by studying the
gas kinematics, revealing a more
complex region than previously thought. Their observations indicated that, in addition
to simple gas photoionization by radiation from the central source, the interaction 
between nuclear mass ejections and /or winds and the interstellar medium play a fundamental role in the morphology and kinematics of the CLR. The analysis of CL profiles showed clear signatures of bipolar movements, associated with radio jets, and suggested that the CLR is formed both by direct ionization from the central source and through mechanical processes, such as shocks.
Their observations point out the importance of multiple mechanisms that affect
the dynamics of ionized gas in the innermost regions of AGNs. In addition, all the above works
demonstrate that the CLR can be used to trace the presence and
the effects of outflows, which are indicative of the influence of the AGN on scales of hundreds of parsecs, confirming previous findings on this matter.

In summary, in the last 20 years, we have witnessed a redefinition of the so-called CLR, which nowadays is understood as a two component region. The unresolved portion, located in the innermost part of the NLR ($< 30$ pc), and the extended one, at scales of hundreds or even a few thousand of parsecs. This new paradigm cemented the strong interest on these lines, as will be seen in the next section.

\section{Importance of the study of Coronal Lines} 
\label{sec:importance}

Since the late nineties, CLs were pointed out as a suitable tool to trace the ionizing continuum of AGN as the IPs of these lines (IP$>$100~eV) are located in the unobserved and/or strongly absorbed portion of the spectral energy distribution emitted by the central engine. \citet{prieto+00}, for instance, employed the [O\,{\sc iv}]~25.9$\mu$m, [Ne\,{\sc v}]~14.322~$\mu$m, [Mg\,{\sc viii}]~3.027$\mu$m, and [Si\,{\sc ix}]~2.583$\mu$m lines to determine their relationship with the soft part of the ionizing spectrum from 50 to 300~eV. For  NGC\,1068, Circinus, and NGC\,4151, the brightest objects of their sample, a blackbody UV continuum is favored. Nearly simultaneously, \citet{alexander+99} and \citet{alexander+00} through a compilation of UV to NIR narrow emission line data in combination to ISO-SWS~4 2.5-45~$\mu$m infrared spectroscopic observations of the Seyfert galaxies NGC\,4151 and NGC\,1068, reconstructed the intrinsic SED of these two AGN.  Their results were consistent with the picture that luminous Seyfert galaxies are powered by a thin accretion disk that produce a quasi-thermal Big Blue Bump, and that the NLR sees a partially absorbed ionizing continuum.

The relationship between the central source ionizing continuum and CLs was further reinforced after the work of \citet{cann+18}. They showed, by means of theoretical modelling, that when the black hole mass decreases, the hardening of the spectral energy distribution of the accretion disk causes infrared coronal lines with the highest ionization potentials to become prominent, revealing a powerful probe of black hole mass in AGNs. That prediction was brought a steep further by \citet{prieto+22}, who proposed the first relationship between the black hole mass and CL emission through the flux ratio [Si\,{\sc vi}]~1.963$\mu$m/Br$\gamma$, being Br$\gamma$ the flux of that line emitted by the BLR. The scatter of the correlation is similar to that of the well-known $M_{\rm BH} - \sigma$ relation \citep{ferrarese_2000}. In addition, the relationship of \citet{prieto+22} confirms that the CL emission produced in the central few parsecs of an AGN is primarily produced by continuum photoionization produced by a geometrically thin optically thick accretion disk. 

Furthermore, CLs play an essential role in studying the evolution of
galaxies hosting AGNs, mainly due to their association with energetic outflows \citep{ardila_2006}.
While [O\,{\sc iii}]~$\lambda$5007 is usually employed as a proxy to trace the ionised phase of outflows \citep{greene_2011}, CLs such as [Fe\,{\sc vii}] and [Ne\,{\sc v}]
offer a complementary and, often, a cleanest view of such processes. This is because of their
high ionization potentials, making that emission free of any contribution due to stellar outflows. Thus, CLs are capable of mapping outflow regions that are fully due to the AGN.  \citep{muller_2011,ardila_2017_ngc_1386,may_2018,fonseca-faria+23}. This is particularly relevant in the extended CLR, where CLs can track outflows like those produced by the interaction between the radio jet and the ISM gas.

Last but not the least, \citet{trindade_2022} showed via photoionisation modeling that CLs from ions with ionization potential greater than or equal to that of O\,{\sc vii}, i.e., 138 eV, trace the footprint of X-ray gas. Thus, they can be used to measure the kinematics of the nuclear and the extended X-ray emitting gas at the spectral resolution dictated by optical spectrographs. In this respect, coronal lines allow us to detect the high-, and the highest-excitation component of the outflow in AGN.

In this way, CLs not only reveal the physics of the central regions of AGNs,
but they also help us to understand how SMBHs impact their galaxies
hosts throughout cosmic time. The detailed study of outflows with
CLs offer an unprecedented opportunity to map a component that is not detectable with other emission lines, sheading light to new aspects of galaxy evolution
and the interaction between AGNs and their host galaxies.  

\section{Open questions and the Future of the study of CLs} 
\label{sec:open}

The interest in the study of the coronal lines has expanded considerably due to their strong relationship with the central ionising continuum and black hole mass. Moreover, the realization that the extended coronal emission is mostly shock-driven, likely powered by the interaction of the jet and the ISM open new perspectives to the study of the ionised phase of the feedback in AGNs.  Still, after decades of intensive research, many open questions, some of them related to these topics, are currently the subject of intensive scrutiny. In all cases, the main goal is to expand our knowledge of the physical mechanisms leading to that emission and the additional information it can bring to our understanding of the AGN phenomenon. 

While [Si\,{\sc vi}] offers a versatile tool for determining the BH mass in Type~I AGN, there are two limitations: the line is close to a strong H$_2$O telluric band and it is out of the $K$-band spectral coverage for sources with redshifts larger than $z$ = 0.22. Therefore, it is necessary to search for CLs of similar ionisation potentials that correlate with the BH mass and are accessible in sources at larger redshifts. Fortunately, the JWST has opened up a new window for probing near-infrared coronal lines previously difficult to access from the ground due to atmospheric absorption around the 3$\mu$m spectral region. Such is the case of [Si\,{\sc vii}]~2.48$\mu$m (IP =205 eV) and [Mg\,{\sc viii}]~3.03$\mu$m (IP = 224 eV). They both  reside in relatively spectrally clean regions, offering a robust tool for studying AGN environments across all demographics and cosmic times.

A key application for coronal lines in resolved studies of AGN-driven outflows is the computation of the outflowing gas masses, energetics, and physical extents \citep{ardila_2017_ngc_1386,may_2018,muller+18,ardila_2020,fonseca-faria+23}. Particularly, close to the base of the outflow in the AGN vicinity, the dust geometry and dense gas density, both of which influence the illumination of the wind tracers, may obscure or bias the intrinsic morphology and kinematics of the outflows. Having proper measurements of outflow properties for sources spanning a range of AGN luminosities and torus viewing angles will help differentiate between various models of multiphase AGN feedback at small physical scales \citep{richardson+16}. Thus, sensitive, spatially and spectrally resolved observations are crucial for constraining the small-scale properties of outflows commonplace within active galaxies.

A search in the literature shows that studies published before the 2020's identifying CLs in AGN outside the local universe ($z > 0.1$)  are scarce \citep{gilli+10,ardila_2011,mignoli+13,rose+15a,lamperti+17}. Among them, \citet{gilli+10} and \citet{mignoli+13} deserve special attention because they employed the CL of [Ne\,{\sc v}]~$\lambda$3426 to detect obscured AGN from optical spectroscopic surveys based on the presence of that line. That technique continues to be employed in more recent surveys. For example,  \citet{cleri+23} analyse a sample of 25 [Ne\,{\sc v}]~$\lambda$3426 emission-line galaxies at 1.4 $< z <$ 2.3 using the Hubble Space Telescope/Wide Field Camera~3 and grism observations.  In combination with lines of lower IP such as [O\,{\sc iii}], [Ne\,{\sc iii}], [O\,{\sc ii}] and [S\,{\sc ii}], they cover a large range of ionization states, allowing to trace multiple phases in the ISM.  Understanding the population of high-ionization galaxies is fundamental for studies of the epoch of reionization. In this respect, future works with JWST  will allow to uncover these extreme [Ne\,{\sc v}]-emitting galaxies at
0.8 $< z <$ 14 and to study the underlying physics of these  system, which remain poorly understood. Indeed, the recent detection of [Ne\,{\sc v}] emission from a faint epoch of reionization-era galaxy at $z$ = 5.59 \citep{chisholm+24} demonstrates the usefulness of CLs to probe the formation and growth of the first black holes in the universe.

With no doubt, all the above studies need to be accompanied by precise modelling to answer one of the key question involving CLs but still awaiting to be answered. Why not all AGN display these lines and/or why they are so faint in some galaxies? Studies made with samples encompassing a few hundred sources \citep{lamperti+17,denbrok+22} show that between 45-60\% of AGN display at least one coronal line. If the SED of these objects typically extends to hundreds of
electron volts and above, which should be able to produce such highly ionized gas, why a significant fraction of the AGN population do not display them? The work of \citet{mckaig+24} shed some light to this issue by proposing that the lack of optical CL emission can be due to the presence of dust, which would reduce the strength of most CLs by $\sim$3 orders of magnitude, primarily as a result of depletion of metals onto the dust grains. In this scenario,  prominent CL emission likely originates in dustless gas. Their results is supported by theoretical modelling that calculates CL luminosities and equivalent widths from radiation-pressure-confined photoionized gas slabs exposed to an AGN continuum, confirming earlier claims of Binette (1998).  In order to fully validate this scenario, surveys that include thousands of AGN at both optical and NIR and at very different redshifts need to be explored and confronted to the models.

\section{Final Remarks} 
\label{sec:remarks}

In this review we have highlighted the most relevant aspects of the coronal lines in active galactic nuclei, from their early conceptions to the most recent developments and results gathered using JWST. The results show that the study of this emission has gained momentum due to its strong relationship with the ionised component of the mechanical feedback and the possibility to employ it to measure the mass of the central black hole using single-epoch spectroscopy and uncover obscured AGN. Moreover, the detection of CLs at high-redshifts by means of HST and JWST has opened a new perspective at allowing to probe the physics of galaxies in the epoch of reionization. Still, the complete understanding of why CLs are not detected in all AGN remains one of the most important issues to be solved in order to fully understand the coronal spectrum in AGN.

\section*{Conflict of Interest Statement}
%All financial, commercial or other relationships that might be perceived by the academic community as representing a potential conflict of interest must be disclosed. If no such relationship exists, authors will be asked to confirm the following statement: 

The authors declare that the research was conducted in the absence of any commercial or financial relationships that could be construed as a potential conflict of interest.

\section*{Author Contributions}

A.R.A: Writing - Conceptualization, original draft, review \& editing.\\
F.C.C.C: Writing – review \& editing.

%The Author Contributions section is mandatory for all articles, including articles by sole authors. If an appropriate statement is not provided on submission, a standard one will be inserted during the production process. The Author Contributions statement must describe the contributions of individual authors referred to by their initials and, in doing so, all authors agree to be accountable for the content of the work. Please see  \href{https://www.frontiersin.org/about/policies-and-publication-ethics#AuthorshipAuthorResponsibilities}{here} for full authorship criteria.

%\section*{Funding}
%Details of all funding sources should be provided, including grant numbers if applicable. Please ensure to add all necessary funding information, as after publication this is no longer possible.

\section*{Acknowledgments}
A.R.A acknowledges Conselho Nacional de Desenvolvimento Científico e Tecnológico (CNPq) for partial support to this work under grant 313739/2023-4. We thank to the anonymous Referee for useful comments/suggestions that improved this manuscript.

%\section*{Supplemental Data}
% \href{http://home.frontiersin.org/about/author-guidelines#SupplementaryMaterial}{Supplementary Material} should be uploaded separately on submission, if there are Supplementary Figures, please include the caption in the same file as the figure. LaTeX Supplementary Material templates can be found in the Frontiers LaTeX folder.

%\section*{Data Availability Statement}
%The datasets [GENERATED/ANALYZED] for this study can be found in the [NAME OF REPOSITORY] [LINK].
% Please see the availability of data guidelines for more information, at https://www.frontiersin.org/about/author-guidelines#AvailabilityofData

\bibliographystyle{Frontiers-Harvard} %  Many Frontiers journals use the Harvard referencing system (Author-date), to find the style and resources for the journal you are submitting to: https://zendesk.frontiersin.org/hc/en-us/articles/360017860337-Frontiers-Reference-Styles-by-Journal. For Humanities and Social Sciences articles please include page numbers in the in-text citations 
\bibliography{referencia}

\end{document}